\documentclass[final,authoryear,12pt,5p,times]{elsarticle}

\usepackage{bm}
\newcommand{\bmath}[1]{\bm{#1}}

\newcommand{\mathbmss}[1]{\bm{\mathsf{#1}}}

\newcommand{\sign}{\mathop{\text{sign}}\nolimits}
 \usepackage{graphicx}
\usepackage{amssymb}
\usepackage{amsmath}
\journal{Astronomy and Computing}

\begin{document}
\sloppy

\begin{frontmatter}


% use the tnoteref command within \title for footnotes;
% use the tnotetext command for the associated footnote;
% use the fnref command within \author or \address for footnotes;
% use the fntext command for the associated footnote;
% use the corref command within \author for corresponding author footnotes;
% use the cortext command for the associated footnote;
% \author{Name\corref{cor1}\fnref{label2}}
% \address{Address\fnref{label3}}

\title{Fast error--controlling MOID computation for confocal elliptic orbits}

% \author[label1,label2]{<author name>}
% \address[label1]{<address>}
% \address[label2]{<address>}

\author{Roman V. Baluev}
\address{Saint Petersburg State University, Faculty of Mathematics and Mechanics, Universitetskij
pr. 28, Petrodvorets, Saint Petersburg 198504, Russia}
\address{Central Astronomical Observatory at Pulkovo of the Russian Academy of Sciences,
Pulkovskoje sh. 65/1, Saint Petersburg 196140, Russia}
 \ead{r.baluev@spbu.ru}

\author{Denis V. Mikryukov}
\address{Saint Petersburg State University, Faculty of Mathematics and Mechanics, Universitetskij
pr. 28, Petrodvorets, Saint Petersburg 198504, Russia}

\begin{abstract}
We present an algorithm to compute the minimum orbital intersection distance (MOID), or
global minimum of the distance between the points lying on two Keplerian ellipses. This is
achieved by finding all stationary points of the distance function, based on solving an
algebraic polynomial equation of $16$th degree. The algorithm tracks numerical errors
appearing on the way, and treats carefully nearly degenerate cases, including practical
cases with almost circular and almost coplanar orbits. Benchmarks confirm its high numeric
reliability and accuracy, and that regardless of its error--controlling overheads, this
algorithm pretends to be one of the fastest MOID computation methods available to date, so
it may be useful in processing large catalogs.
\end{abstract}


\end{frontmatter}

% \linenumbers


\section{Introduction}
The MOID parameter, or the minimum distance between points on two Keplerian orbits, has an
important value in various Solar System studies. It measures the closeness of two
trajectories in the $\mathbb R^3$ space, and hence ascertains whether two bodies have a
risk to collide. For example, if MOID appeared below the sum of radii of two bodies than
such bodies may avoid a collision only if they orbit in a mean-motion resonance, or via a
perturbating effect that may increase the MOID to a safe level before the bodies could
actually collide. Otherwise, the bodies will necessarily collide in some future.

Therefore, computing the MOID is very old task with an application to Potentially Hazardous
Objects (PHOs) and Near-Earth Asteroids (NEAs).

This problem is investigated over decades already, see e.g. \citep{Sitarski68,Dybczynski86}
and more recent works by \citet{Armellin10,Hedo18}.

The MOID is a minimum of some distance or distance-like function $\rho(u,u')$ that depends
on two arguments, determining positions on two orbits. The methods of finding the minima of
$\rho(u,u')$ can be split in several general cathegories, depending on the dimensionality
of the optimization task to be solved. This depends on how much work is pre-computed
analytically.

\begin{enumerate}
\item Global optimization in 2D. As an ultimately simple example this includes e.g. the 2D
brute-force (exhaustive) search of $\rho(u,u')$ on a 2D grid. Thanks to the existence of
rigorous and finite upper limits on the gradient of $\rho(u,u')$, which appears to be a
trigonometric polynomial, we can always limit the finite difference $\Delta\rho$ by $\Delta
u \max |\rho'_u|$ and $\Delta u' \max |\rho'_{u'}|$. Thanks to such error predictability,
algorithms of the 2D class appear the most reliable ones, because we can always determine
the MOID, along with the orbital positions $u$ and $u'$, to any desired accuracy. An
advanced method based on the 2D global optimization of $\rho$, which includes high-order
Taylor models and interval arithmetics, was presented by \citet{Armellin10}. Nevertheless,
even advanced methods of this type cannot be fast due to the need of considering 2D
domains.

\item 1D optimization. Here we eliminate $u'$ from the numeric search by solving it from an
analytic equation. The remaining orbital position $u$ is determined by numeric optimization
of the 1D function $\tilde\rho(u) = \rho(u,u'(u))$. In general, this is faster than 2D
minimization, but the derivative of $\tilde\rho(u)$ is no longer bounded, because $du'/du$
may turn infinite sometimes. Such cases may appear even for very simple circular orbits.
Therefore, in general this method cannot provide a strict mathematical guarantee of the
desired numerical accuracy. However, it appears more reliable than the methods of the next
class. The SDG method discussed by \citet{Hedo18} basically belongs to this class.

\item Methods of the 0D class, in which both $u$ and $u'$ are solved for rather than found
by numeric optimization. This includes the method by \citet{KholshVas99} and by
\citet{Gronchi02,Gronchi05}, because they do not explicitly deal with any numeric
optimization at all. The task is analytically reduced to solving a nonlinear equation with
respect to $u$ and then express $u'$ also analytically. Methods of this class are
ultimately fast but relatively vulnerable to loosing roots due to numerical errors (in
nearly degenerate cases). This effect becomes important because the equation for $u$ is
quite complicated and subject to round-off errors. Also, this equation often have close
(almost multiple) roots that are always difficult for numeric processing.
\end{enumerate}

Here we present an efficient numeric implementation of the algebraic approach presented by
\citet{KholshVas99}, similar to the one presented by \citet{Gronchi02,Gronchi05}. This
method belongs to the fast 0D class. It is based on analytic determination of all the
critical points of the distance function, $|\bmath r - \bmath r'|^2$, augmented with an
algebraic elimination of one of the two positional variables. Mathematically, the problem
is reduced to a single polynomial equation of $16$th degree with respect to one of the
eccentric anomalies.

Recently, \citet{Hedo18} suggested a method that does not rely on necessary determination
of all the stationary points for the distance. It basically splits the problem in two tasks
of 1D optimization, so this method belongs to the 1D class. Nevertheless, it proved $\sim
20$ per cent faster than the 0D Gronchi algorithm, according to the benchmarks. Though the
performance differences appeared relatively moderate, there were revealed occurences when
the Gronchi's code suffered from numeric errors, reporting a wrong value for the MOID.
Therefore, in this task the numeric reliability of the method is no less important than
just the computing speed.

Direct implementation of the methods by \citet{KholshVas99} and \citet{Gronchi02} might be
vulnerable, because finding roots of a high-degree polynomial might be a numerical
challenge sometimes. When dealing with large asteroid catalogs, various almost-degenerate
cases appear sometimes, if the equations to be solved contain almost-double or
almost-multiple roots. Such roots are difficult to be estimated accurately, because they
are sensitive to numeric errors (even if there were no errors in the input orbital
elements). Moreover, we have a risk of ambiguity: if the polynomial has two or more very
close real roots then numeric errors may result in moving them to the complex plane
entirely, so that we may wrongly conclude that there are no such real roots at all. Such
effect of lost of real roots may potentially result in overestimating the MOID, i.e. it may
appear that we lost exactly the solution corresponding to the global minimum of the
distance. This issue can be solved by paying attention not only to the roots formally
identified as real, but also to all complex-valued roots that appear suspiciously close to
the real axis. To define formally what means `suspiciously close' we need to estimate
numeric error attached to a given root, not just its formal value.

In other words, our task assignes an increased role to the numeric stability of the
computation, because errors are known to dramatically increase when propagating through
mathematical degeneracies. This motivated us to pay major attention to error control when
implementing the method by \citet{KholshVas99} in a numerical algorithm.

The structure of the paper is as follows. In Sect.~\ref{sec_math}, we give some
mathematical framework that our algorithm relies upon. Sect.~\ref{sec_alg} describes the
numeric algorithm itself. Sect.~\ref{sec_tols} contains some guidelines on how to select
meaningful error tolerances for our algorithm. Sect.~\ref{sec_perf} presents its
performance tests. In Sect.~\ref{sec_add}, we describe several auxiliary tools included in
our MOID library.

The C++ source code of our MOID library named {\sc distlink} is available for download at
\texttt{http://sourceforge.net/projects/distlink}.

\section{Mathematical setting}
\label{sec_math}
Consider two confocal elliptic orbits: $\mathcal E$ determined by the five geometric
Keplerian elements $a,e,i,\Omega,\omega$, and $\mathcal E'$ determined analogously by the
same variables with a stroke. Our final task is to find the minimum of the distance
$|\bmath r - \bmath r'|$ between two points lying on the corresponding orbits, and the
orbital positions $u,u'$ where this minimum is attained (here $u$ stands for the eccentric
anomaly). According to \citet{KholshVas99}, this problem is reduced to solving for the
roots of a trigonometric polynomial $g(u)$ of minimum possible algebraic degree $16$
(trigonometric degree $8$). It is expressed in the following form:
\begin{align}
g(u) &= K^2 (A^2-C^2) (B^2-C^2) + \nonumber\\
&+ 2 K C \left[NA (A^2-C^2) + MB (B^2-C^2)\right] - \nonumber\\
&- (A^2+B^2) \left[N^2(A^2-C^2)+M^2(B^2-C^2)-\right.\nonumber\\
&\left.\phantom{(A^2+B^2)}-2NMAB\right],
\label{gdef}
\end{align}
where
\begin{eqnarray}
A &=& PS' \sin u - SS' \cos u, \nonumber\\
B &=& PP' \sin u - SP' \cos u, \nonumber\\
C &=& e' B - \alpha e \sin u (1-e\cos u), \nonumber\\
M &=& PP' \cos u + SP' \sin u + \alpha e' - PP' e, \nonumber\\
N &=& PS' e - SS' \sin u - PS' \cos u, \nonumber\\
K &=& \alpha' e'^2,
\label{ABC}
\end{eqnarray}
and $\alpha=a/a'$, $\alpha'=a'/a$. The quantities $PP'$, $PS'$, $SP'$, $SS'$ represent
pairwise scalar products of the vectors $\bmath P$ and $\bmath S$:
\begin{align}
\bmath P = \{&\cos\omega\cos\Omega-\cos i\sin\omega\sin\Omega, \nonumber\\
&\cos\omega\sin\Omega+\cos i\sin\omega\cos\Omega, \nonumber\\
&\sin i\sin\omega\, \}, \nonumber\\
\bmath S = \phantom{\{} &\bmath Q \sqrt{1-e^2}, \nonumber\\
\bmath Q = \{&-\sin\omega\cos\Omega-\cos i\cos\omega\sin\Omega, \nonumber\\
&-\sin\omega\sin\Omega+\cos i\cos\omega\cos\Omega, \nonumber\\
&\sin i\cos\omega\, \},
\end{align}
with analogous definitions for $\bmath P'$ and $\bmath S'$.

When all the roots of $g(u)$ are found, for each $u$ we can determine the second position
$u'$ from
\begin{equation}
\cos u' = \frac{BC + mA\sqrt D}{A^2+B^2},\; \sin u' = \frac{AC - mB\sqrt D}{A^2+B^2},
\label{us}
\end{equation}
where
\begin{equation}
D = A^2+B^2-C^2, \quad m=\pm 1.
\end{equation}
The sign of $m$ should be chosen to satisfy
\begin{equation}
M \sin u' + N \cos u' = K \sin u'\cos u',
\end{equation}
so there is only a single value of $u'$ that corresponds to a particular solution for $u$.

Finally, after both the orbital positions $u$ and $u'$ were determined, the squared
distance between these points is $|\bmath r - \bmath r'|^2 = 2aa'\rho(u,u')$, where
\begin{align}
\rho(u,u') = &\frac{\alpha+\alpha'}{2}+\frac{\alpha e^2+\alpha' e'^2}{4} - PP'ee' + \nonumber\\
&+ (PP'e'-\alpha e)\cos u + SP'e'\sin u + \nonumber\\
&+ (PP'e-\alpha'e')\cos u'+PS'e\sin u'- \nonumber\\
&- PP'\cos u\cos u' - PS'\cos u\sin u'- \nonumber\\
&- SP'\sin u\cos u' - SS'\sin u\sin u' + \nonumber\\
&+ \frac{\alpha e^2}{4}\cos 2u + \frac{\alpha'e'^2}{4} \cos 2u'.
\label{rho}
\end{align}

Therefore, our general computation scheme should look as follows: (i) find all real roots
of $g(u)$; (ii) for each solution of $u$ determine its corresponding $u'$; (iii) for each
such pair $u,u'$ compute $\rho(u,u')$; and (iv) among these values of $\rho$ select the
minimum one. This will give us the required MOID estimate. As we can see, the most
difficult step is finding all real roots of the trigonometric polynomial $g(u)$, while the
rest of the work is rather straightforward.

This trigonometric polynomial can be rewritten in one of the two standard forms:
\begin{equation}
g(u) = a_0 + 2 \sum_{k=1}^N (a_k \cos ku + b_k \sin ku) = \sum_{k=-N}^N c_k {\mathrm e}^{iku},
\label{gcan}
\end{equation}
where $N=8$. The coefficients $a_k$, $b_k$, and $c_{\pm k} = a_k\mp i b_k$ can be expressed
as functions of the quantities $PP'$, $PS'$, $SP'$, $SS'$, and $\alpha$, $e$, $e'$. Most of
such explicit formulae would be too huge and thus impractical, but nonetheless we computed
an explicit form for the coefficient $c_8$:
\begin{align}
c_8 = c_{-8}^* &= \left(\frac{\alpha e^2}{16}\right)^2 M_1 M_2 M_3 M_4, \nonumber\\
M_1 &= PP'-SS' - ee' - i (SP'+PS'), \nonumber\\
M_2 &= PP'-SS' + ee' - i (SP'+PS'), \nonumber\\
M_3 &= PP'+SS' - ee' - i (SP'-PS'), \nonumber\\
M_4 &= PP'+SS' + ee' - i (SP'-PS').
\label{c8}
\end{align}
Here the asterisk means complex conjugation.

The number of real roots of $g(u)$ cannot be smaller than $4$ \citep{KholshVas99}. Also,
this number is necessarily even, since $g(u)$ is continuous and periodic. But the upper
limit on the number of real roots is uncertain. In any case, it cannot exceed $16$, the
algebraic degree of $g(u)$, but numerical simulations performed by \citet{KholshVas99}
never revealed more than $12$ real roots. Here we reproduce their empirical upper limit:
based on a test computation of $\sim 10^8$ orbit pairs from the Main Belt (see
Sect.~\ref{sec_perf}), we obtained approximately one $12$-root occurence per $\sim 4\times
10^6$ orbit pairs\footnote{This is actually an upper limit on that rate, because our
algorithm may intentionally count some complex roots with small imaginary part as real
ones. This estimate is sensitive to the selected floating-point precision and to subtle
details that affect overall numeric accuracy of the algorithm. It may even be possible that
all these potential $12$-root occurences contain only $10$ real roots.}. No cases with $14$
or $16$ roots were met.\footnote{We had one $14$-root occurence using the standard {\sc
double} precision, but this case appeared to have only $12$ real roots with {\sc long
double} arithmetic.}

Since the number of real roots of $g(u)$ is highly variable and a priori unknown, certain
difficulties appear when dealing with $g(u)$ in the real space. In practice $g(u)$ often
becomes close to being degenerate, e.g. in the case of almost circular or almost coplanar
orbits, which is frequent for asteroids and typical for major planets in the Solar System.
In such cases, real roots of $g(u)$ combine in close pairs or even close quadruples. The
graph of $g(u)$ passes then close to the abscissa near such roots. This means that numeric
computing errors affect such nearly-multiple roots considerably, implying increased
uncertainties. Moreover, we might be even uncertain about the very existence of some roots:
does the graph of $g(u)$ really intersects the abscissa or it passes slightly away, just
nearly touching it? In practice this question may become non-trivial due to numerical
errors, that might appear important because $g(u)$ is mathematically complicated.
Therefore, treating $g(u)$ only in the real space might result in loosing its roots due to
numeric errors.

But loosing real roots of $g(u)$ would potentially mean to vulnerably overestimate the
MOID, because there is a risk that the minimum distance $|\bmath r-\bmath r'|$
occasionally corresponds to a lost root. Then it might be more safe to overestimate the
number of real roots of $g(u)$, i.e. we should also test ``almost-real'' complex roots that
correspond to a near-touching behaviour of $g(u)$, even if it does not apparently intersect
the abscissa. This would imply some computational overheads and additional CPU time
sacrificed for the algorithmic reliability and numeric stability. Also, this would mean to
treat $g(u)$ in the complex plane and find all its complex roots, rather than just the real
ones. As such, we need to swap to complex notations.

By making the substitution $z={\mathrm e}^{iu}$ or $w={\mathrm e}^{-iu}$, we can transform
$g(u)$ to an algebraic polynomial of degree $16$:
\begin{equation}
g(u) = \sum_{k=-N}^N c_k z^k = \mathcal P(z) w^N = \mathcal Q(w) z^N.
\label{gPQ}
\end{equation}
So, the task of finding roots of $g(u)$ becomes equivalent to solving $\mathcal P(z)=0$ or
$\mathcal Q(w)=0$.\footnote{Since $u$ can take only real values, we always have $z\neq 0$
and $w\neq 0$.} Among all these complex roots we must select those that within numeric
errors lie on the unit circle $|z|=|w|=1$.

Since all $a_k$ and $b_k$ are real, the complex coefficients satisfy the property $c_k =
c_{-k}^*$. Hence, roots of $\mathcal P(z)$ obey the following rule: if $z=r{\mathrm
e}^{i\varphi}$ is such a root then $1/z^*=r^{-1}{\mathrm e}^{i\varphi}$ is also a root of
$\mathcal P$. Therefore, the entire set of these roots includes three families: (i) roots
on the unit circle $|z|=1$ that correspond to real $u$, (ii) roots outside of this circle,
$|z|>1$, and (iii) roots inside the unit circle, $|z|<1$. The roots with $|z|\neq 1$ are
split into pairs of mutually inverse values that have $|z|<1$ and $|z|>1$.

\section{Numerical algorithm}
\label{sec_alg}
\subsection{Determining the polynomial coefficients and their uncertainty}
First of all, we must represent the polynomial $g(u)$ in its canonical form~(\ref{gcan}).
For that, we need to compute the coefficients $c_k$. The explicit formulae for $c_k$ are
too complicated and impractical, except for the case $k=\pm 8$ given in~(\ref{c8}). Instead
of direct computation of $c_k$, we determine them by means of the discrete Fourier
transform (DFT hereafter):
\begin{equation}
c_k = \frac{1}{2N+1} \sum_{m=0}^{2N} g(u_m)\, {\rm e}^{ik u_m}, \quad u_m=\frac{2\pi m}{2N+1}.
\label{dft}
\end{equation}
Here, $g(u_m)$ are computed by using the relatively compact formula~(\ref{gdef}). Regardless
of the use of DFT, this approach appears computationally faster than computing all $c_k$
directly. We do not even use FFT algorithms for that, because of too small number of
coefficients $N=8$. For so small $N$, the FFT technique did not give us any remarkable
speed advantage in comparison with the direct application of the DFT~(\ref{dft}).

However, the DFT can likely accumulate some rounding errors. The accuracy of so-determined
$c_k$ can be roughly estimated by comparing the DFT estimate of $c_{\pm 8}$ with its
explicit representation~(\ref{c8}), which is still mathematically simple. We may assume
that numerical errors inferred by the formula~(\ref{c8}) are negligible, and that all the
difference between~(\ref{dft}) and~(\ref{c8}) is solely explained by the DFT errors.

Moreover, we can compute the DFT~(\ref{dft}) for any $N>8$. In such a case, all the
coefficients $c_k$ for $|k|>8$ must turn zero. However, due to numeric errors their DFT
estimate may occur non-zero, and in such a case the magnitude of this $c_k$ can be used as
a rough error assessment.

Based on this consideration, we adopted the following formulae to estimate the average
error in $c_k$:
\begin{equation}
\varepsilon^2 = \left.\left(|c_8-c_8'|^2 + \sum_{k=9}^N |c_k'|^2 \right)\right/(N-7).
\label{err}
\end{equation}
Here, $c_8$ is determined from~(\ref{c8}), while $c_k'$ are DFT estimations
from~(\ref{dft}). The formula~(\ref{err}) represents a statistical estimation that treats
numerical errors in $c_k$ as random quantities. It is based on the assumption that errors
in different $c_k$ are statistically independent (uncorrelated) and have the same variance.
In such a case, $\varepsilon^2$ provides an estimate of that variance.

In our algorithm, we set $N=10$, thus computing the DFT from $21$ points $u_m$. In
practical computation we always obtained $\varepsilon$ not far from the machine precision,
except for rare cases. We additionally notice that the error estimation~(\ref{err}) also
includes, to a certain extent at least, the numeric error appeared when computing the
values of $g(u_m)$ by formula~(\ref{gdef}), not just the DFT errors inferred
by~(\ref{dft}).

\subsection{Root-finding in the complex plane}
When all $c_k$ are determined, along with their probable numerical error, we can determine
all complex roots of $\mathcal P(z)$. This is done via Newtonian iterations and obeys the
following numeric scheme:
\begin{enumerate}
\item Initial approximations for the first $8$ roots are selected in a specific optimized
manner as detailed below.
\item Initial approximation for each next root $z_k$ is chosen according to the prediction
$z_k^{(0)}=1/z_{k-1}^*$, where $z_{k-1}$ is the final estimation of the previous root.
Thanks to such a choice, the algorithm will always extract a paired complex root
$z_k=1/z_{k-1}^*$ immediately after $z_{k-1}$. The Newtonian iterations for $z_k$ converge
in this case very quickly (in two iterations or so). This does not work, if $z_{k-1}$
belongs to the family $|z|=1$ (such roots do not combine into inverse pairs), or if
$z_{k-1}$ turns out to be that \emph{second} root in the pair. Then such a starting
approximation would be equal to either $z_{k-1}$ or $z_{k-2}$, so the next extracted root
$z_k$ will likely appear close to one of these.
\item Each root is refined by Newtonian iterations (i) until some least
required relative accuracy $\delta_{\max}$ is achieved, and then (ii) until we reach the
desired target relative accuracy $\delta_{\min}$ or, at least, the maximum possible machine
accuracy, if $\delta_{\min}$ is unreachable. On the first phase, we iterate until the last
Newtonian step $|d_n|$ falls below $\delta_{\max}|z|$. The iterations are restarted from a
different starting point, if they are trapped in an infinite loop at this phase (this is
the known curse of the Newton method). On the second phase, the stopping criterion relies
on the last and pre-last Newtonian steps, $|d_n|$ and $|d_{n-1}|$. The iterations are
continued either until $|d_n|<\delta_{\min}|z|$, or until the relative step change,
$\gamma_n=(|d_{n-1}|^2-|d_n|^2)/|d_n|^2$, drops below the machine epsilon $\epsilon$. The
latter criterion is motivated as follows. In the middle of iterations, whenever numeric
round-off errors are not significant yet, the parameter $\gamma_n$ should remain large
positive, since each $|d_n|$ is much smaller than $|d_{n-1}|$. But in the end either
$\gamma_n\to 0$, if iterations get finally stuck at almost the same numeric value near the
root, or $\gamma_n$ occasionally attains negative values, if the iterations start to
randomly jump about the root due to numeric errors. A good practical assumption for the
accuracy parameters might be $\delta_{\max}\sim \sqrt{\epsilon}$ and $\delta_{\min}=0$ or
about $\epsilon$.
\item Whenever we have an accurate estimate of a root $z_k$, this root is eliminated from
$\mathcal P(z)$ through dividing it by $(z-z_k)$ via the Horner scheme. The remaining
polynomial has a reduced degree. For the sake of numerical stability, we either extract the
multiplier $(z-z_k)$ from $\mathcal P(z)$, if $|z_k|>1$, or $(w-w_k)$ from $Q(w)$, if
$|z_k|<1$.
\item The roots are extracted in such a way until we receive a quadratic polynomial in
$\mathcal P(z)$. Its two remaining roots are then obtained analytically.
\end{enumerate}

The order, in which the roots are extracted, is important. If we extract `easy' roots
first, we spend little Newtonian iterations with a high-degree $\mathcal P$. Also, such
`easy' roots should likely be far from degeneracies and hence be numerically accurate.
Therefore, they should not introduce big numeric errors when the Horner scheme is applied.
The `difficult' roots that require big number of Newtonian iterations should better be
extracted later, when the degree of $\mathcal P$ is reduced. If we act in an opposite
manner, i.e. extract `difficult' roots first, these difficult roots will inevitably
increase numeric errors. After applying the Horner scheme, these increased errors are
transferred to the coefficients $c_k$, reducing the accuracy of all the remaining roots.
Also, bad roots always require larger number of Newtonian iterations, which become even
more expensive at the beginning, when the degree of $\mathcal P$ is still large and its
computation is more slow.

After some tests we decided that the best way is to extract in the very beginning extreme
complex roots: $|z|\ll 1$ and their inversions $|z|\gg 1$. Such roots are determined
quickly and accurately, and the Horner scheme is very stable for them. Since in practical
computations we always revealed at least $4$ complex roots, we try to extract these four
roots in the beginning. The starting approximation for the first root, $z_1^{(0)}$, is
always set to zero. This will likely give us the root with smallest $|z_1|$. The next root,
$z_2$, is started from $z_2^{(0)}=1/z_1^*$ and will be determined almost immediately. It
will be the largest one. Initial approximations for the next too roots, $z_3$ and $z_4$,
are set from our usual rule, $z_k^{(0)}=1/z_{k-1}^*$. Thanks to this, we obtain yet another
smallest root as $z_3$, and yet another largest root as $z_4$.

After these four extreme complex roots are removed from $\mathcal P$, we try to extract the
four guaranteed roots that lie on the unit circle. We select their initial approximations
to be such that $u$ is located at the orbital nodes or $\pm 90^\circ$ from them. This is
motivated by the practical observation that the MOID is usually attained near the orbital
nodal line, see Sect.~\ref{sec_add}. Thanks to such a choice, these four roots are
determined in a smaller number of Newtonian iterations.

The ninth root is iterated starting from $z_9^{(0)}=0$ again, and for the rest of roots we
follow the general rule $z_k^{(0)}=1/z_{k-1}^*$. Thanks to such a choice, the algorithm
tries to extract the remaining roots starting far from the unit circle $|z|=1$, approaching
it in the end. Therefore, the most numerically difficult cases, which are usually located
at $|z|=1$, are processed last, when the degree of $\mathcal P$ is already reduced in a
numerically safe manner.

Using this optimized sequence we managed to reduce the average number of Newtonian
iterations from $7$--$8$ per root to $5$--$6$ per root, according to our benchmark test
case (Sect.~\ref{sec_perf}). Also, this allowed to further increase the overall numeric
accuracy of the roots and numeric stability of the results, because highly accurate roots
are extracted first and roots with poor accuracy did not affect them.

\subsection{Estimating roots uncertainty and roots selection}
When all complex roots of $\mathcal P(z)$ are obtained, we need to select those roots that
satisfy $|z|=1$ and thus correspond to real values of $u$. However, in practice the
equation $|z|=1$ will never be satisfied exactly, due to numerical errors. We need to apply
some criterion to decide whether a particular $|z_k|$ is close to unit, within some
admissible numeric errors, or not.

We approximate the relative error of the root $z_k$ by the following formula:
\begin{equation}
\varepsilon_z^2 = \frac{1}{|z_k|^2} \left( |d|^2 + \frac{\varepsilon_{\mathcal P}^2}{|\mathcal D|^2} \right).
\label{zerr}
\end{equation}
Its explanation is as follows.

Firstly, $d$ is the smaller (in absolute value) of the roots of a quadratic polynomial that
approximates $\mathcal P(z)$ near $z_k$:
\begin{align}
\frac{\mathcal P''(z_k)}{2} d^2 + \mathcal P'(z_k) d + \mathcal P(z_k) = 0, \nonumber\\
d=\frac{-\mathcal P' + \mathcal D}{\mathcal P''}, \quad \mathcal D = \pm \sqrt{\mathcal P'^2 - 2\mathcal P \mathcal P''}.
\label{qapp}
\end{align}
Thus, the first term in~(\ref{zerr}), or $|d|$, approximates the residual error of $z_k$
still remained after Newtonian iterations. It is zero if $\mathcal P(z_k)=0$ precisely.
Here we use the initial polynomial $\mathcal P$ of $16$th degree, not the one obtained
after dividing it by any of $z-z_k$. For practical purposes, $d$ should be calculated using
a numerically stabilized formula that avoids subtraction of close numbers whenever
$\mathcal P\approx 0$. For example, we can use
\begin{equation}
d = \frac{-2\mathcal P}{\mathcal P' + \mathcal D},
\end{equation}
selecting such sign of $\mathcal D$ that maximizes the denominator $|\mathcal P' + \mathcal
D|$.

But just $|d|$ is not enough to characterize the uncertainty of $z_k$ in full. In fact,
most of this uncertainty comes from the numerical errors appearing in $\mathcal P(z)$
through $c_k$. Inaccurate computation of $\mathcal P(z)$ leads to errors in the estimated
root $z_k$. Using the quadratic approximation~(\ref{qapp}), the sensitivity of $z_k$ with
respect to varying $\mathcal P$ is expressed by the derivative $\partial d/\partial\mathcal
P = -1/\mathcal D$. Hence, the second error term in~(\ref{zerr}) appears,
$\varepsilon_{\mathcal P}/|\mathcal D|$, where $\varepsilon_{\mathcal P}$ represents the
error estimate of $\mathcal P(z)$:
\begin{equation}
\varepsilon_{\mathcal P}^2 = \varepsilon^2 \sum_{n=0}^{16} |z_k|^{2n},
\end{equation}
where $\varepsilon$ given in~(\ref{err}).

The quadratic approximation~(\ref{qapp}) is related to the iterative Muller method that
takes into account the second derivative of $\mathcal P$. We needed to take into account
$\mathcal P''$ because in practice the real roots of $g(u)$ are often combined into close
pairs, triggering a close-to-degenerate behaviour with small $|\mathcal P'(z)|$. In such a
case the linear (Newtonian) approximation of $\mathcal P(z)$ yields too pessimistic error
estimate for $z_k$. The use of the quadratic approximation~(\ref{qapp}) instead allows us
to adequately treat such cases with nearly double roots.

However, even with~(\ref{qapp}) it is still difficult to treat the cases in which the roots
combine in close quadruples. Then $\mathcal P''(z_k)$ becomes small too, along with
$\mathcal P'(z_k)$ and $\mathcal P(z_k)$. The error estimate~(\ref{zerr}) becomes too
pessimistic again. Such cases are very rare, but still exist. They may need to be processed
with an alternative computation method (see Sect.~\ref{sec_add}).

In the error estimate~(\ref{zerr}), we neglect numerical errors of $\mathcal P'(z_k)$ and
of $\mathcal P''(z_k)$, assuming that these quantities do not vanish in general and thus
always keep a satisfactory relative accuracy (this is typically true even for almost double
paired roots).

We use the following numeric criterion to identify roots lying on the unit circle:
\begin{equation}
\Delta_z = \frac{\left|\log|z| \right|}{\nu \varepsilon_z} \leq 3.
\label{thr}
\end{equation}
Here, $\nu$ is an auxiliary scaling parameter controlling the tolerance of the threshold.
Normally, it should be set to unit and its meaning is to heuristically correct the
estimated $\varepsilon_z$ in case if there are hints that this error estimation is
systematically wrong. The threshold $3$ is supposed to mean the so-called three-sigma rule.
It was selected well above the unit in order to increase the safety of roots selection and
hence the reliability of the entire algorithm.

After selecting all the roots $z_k$ that lie close enough to the unit circle, we may
determine the corresponding eccentric anomaly $u_k=\arg z_k$, then its corresponding $u_k'$
from~(\ref{us}) and $\rho_k = \rho(u_k,u_k')$ from~(\ref{rho}). The minimum among all
computed $\rho_k$ yields us the required MOID estimate.

In general, the discriminant $D$ in~(\ref{us}) is non-negative if $u$ is a root of $g(u)$,
but this can be violated in some special degenerate cases \citep{Baluev05}. Formally,
negative $D$ means that MOID cannot be attained at the given orbital position $u$, even if
$g(u)=0$. This is quite legal, meaning that some roots of $g(u)$ may be parasitic, i.e.
corresponding to a critical point of $\rho(u,u')$ for some complex $u'$ (even if $u$ is
real). However, $D$ may also turn negative due to numeric errors perturbing almost-zero
$D>0$. We could distinguish such cases based on some uncertainty estimate of $D$, but in
practice it appears easier to process all them just forcing $D=0$. In the first case (if
$D$ is negative indeed), this would imply just a negligible computation overhead because of
unnecessary testing of an additional $\rho_k$ that cannot be a MOID. But in the second case
(if $D$ appeared negative due to numeric errors) we avoid loosing a potential MOID
candidate $\rho_k$.

\subsection{Refining the MOID by 2D iterations}
Now we have quite a good approximation for the MOID and for the corresponding positions in
$u$ and $u'$. However, their accuracy is typically 1-2 significant digits worse than the
machine precision (even if we iterated the roots $z_k$ to the machine precision). The loss
of numeric precision appears in multiple places: in rather complicated formulae
like~(\ref{gdef}), in~(\ref{us}) if $D$ appeared small, in the DFT computation~(\ref{dft}),
and so on. As a result, whenever working in the standard {\sc double} precision, we may
receive average numeric errors of $\sim 10^{-14}$ instead of the relevant machine epsilon
$\sim 10^{-16}$. Although this average accuracy is pretty good for the most practical
needs, in poorly-conditioned cases the errors may increase further. But fortunately, the
results can be easily refined to the machine precision $\sim 10^{-16}$ at the cost of
negligible overheads.

This can be achieved by applying the 2D Newton iteration scheme to the function
$\rho(u,u')$. Let us decompose it into the Taylor series:
\begin{equation}
\rho(u,u') = \rho_0 + \bmath g \cdot \bmath d + \frac{1}{2} \bmath d^{\rm T} \mathbmss H \bmath d + \ldots
\label{rhoTayl}
\end{equation}
Here, $\bmath g$ is the gradient and $\mathbmss H$ is the Hessian matrix of $\rho$,
considered at the current point $(u,u')$, while $\bmath d$ is the 2D step in the plane
$(u,u')$. We need to find such $\bmath d$ where the gradient of~(\ref{rhoTayl}) vanishes:
\begin{equation}
\nabla \rho = \bmath g + \mathbmss H \bmath d + \ldots = 0,
\end{equation}
therefore the necessary 2D step is
\begin{equation}
\bmath d = - \mathbmss H^{-1} \bmath g.
\label{step}
\end{equation}

To compute $\rho$, $\bmath g$, and $\mathbmss H$, we do not rely on the
formula~(\ref{rho}), because it is poorly suited for practical numeric computations. It may
generate a precision loss due to the subtraction of large numbers. Such a precision loss
appears when MOID is small compared to $a$ and $a'$.

The derivatives of $\rho$ can be computed using the following formulae, obtained by direct
differentiation:
\begin{align}
\rho = \frac{(\bmath r - \bmath r')^2}{2aa'}, \;
g_u = \frac{(\bmath r - \bmath r') \bmath r_u}{2aa'}, \;
g_{u'} = -\frac{(\bmath r - \bmath r') \bmath r'_{u'}}{2aa'}, \nonumber\\
H_{uu} = \frac{(\bmath r - \bmath r') \bmath r_{uu} + \bmath r_u^2}{2aa'}, \;
H_{u'u'} = \frac{(\bmath r - \bmath r') \bmath r'_{u'u'} + {{\bmath r}'_{u'}}^2}{2aa'}, \nonumber\\
H_{uu'} = -\frac{\bmath r_u \bmath r'_{u'}}{2aa'}, \qquad \bmath r_u = \frac{d\bmath r}{du}, \quad \bmath r'_{u'} = \frac{d\bmath r'}{du'}.
\label{vgH}
\end{align}
In fact, the effect of the precision loss is present in~(\ref{vgH}) too, due to the
difference $\bmath r-\bmath r'$, but formula~(\ref{rho}) would exacerbate it further,
because it hiddenly involves subtraction of \emph{squares} of the quantities.

Now, according to \citet{Baluev05}, the radius-vector $\bmath r$ on a Keplerian elliptic
orbit is
\begin{equation}
\frac{\bmath r}{a} = \bmath P (\cos u - e) + \bmath S \sin u,
\label{rvec}
\end{equation}
with a similar expression for $\bmath r'$. The corresponding derivatives with respect to
$u$ and $u'$ are obvious.

The stopping criterion for the 2D iterations~(\ref{step}) is similar to the one used in the
Newton-Raphson scheme for the roots $z_k$. It applies the tolerance parameter
$\delta_{\min}$ to $|\bmath d|$. The iterations are therefore continued either until this
accuracy $\delta_{\min}$ is reached by the angular variables $u$ and $u'$, or until we
reach the maximum possible numeric precision, so that further iterations unable to increase
it. The other control parameter $\delta_{\max}$ is not used here.

In practice it appears enough to make just one or two refining iterations~(\ref{step}) to
reach almost the machine accuracy in $\rho$. In rare almost-degenerate cases we may need
$n=3$ iterations or more, but the fraction of such occurences is small and quickly
decreases for larger $n$.

\subsection{Estimating uncertainties of the MOID and of its orbital positions}
The numeric errors in $u$ and $u'$ come from three sources: the floating-point `storage'
errors, the residual errors appearing due to inaccurate fit of the condition
$\nabla\rho=0$, and the numeric errors appearing when computing $\bmath g$. Each of these
error components transfers to $\rho$.

The first error part in $u,u'$ can be roughly approximated as
\begin{equation}
\sigma_{u,u'}^{(1)} = \pi\nu\epsilon,
\label{uu1}
\end{equation}
assuming that $\nu$ is our universal error scaling factor. Since the gradient $\bmath g$ is
negligible near the MOID, the Taylor decomposition~(\ref{rhoTayl}) implies the following
error in $\rho$:
\begin{equation}
\Delta\rho \simeq \frac{1}{2} \bmath d^{\rm T} \mathbmss H \bmath d,
\end{equation}
where $\bmath d$ has the meaning of the 2D numeric error in $u,u'$. From~(\ref{uu1}), we
know only the typical length of $\bmath d$, but the direction of this vector can be
arbitrary. Then we can use the Rayleigh inequality
\begin{equation}
|\bmath d^{\rm T} \mathbmss H \bmath d| \leq |\lambda_{\max}| \bmath d^2,
\end{equation}
where $\lambda_{\max}$ is the maximum (in absolute value) eigenvalue of $\mathbmss H$.
Since the size of $\mathbmss H$ is only $2\times 2$, it can be computed directly:
\begin{equation}
|\lambda_{\max}| = \left|\frac{H_{uu}+H_{u'u'}}{2}\right| + \sqrt{\left(\frac{H_{uu}-H_{u'u'}}{2}\right)^2 + H_{uu'}^2}.
\end{equation}
So, the indicative uncertainty in $\rho$ coming from this error source is
\begin{equation}
\sigma_\rho^{(1)} \simeq \frac{|\lambda_{\max}|}{2} \left(\sigma_u^{(1)}\right)^2.
\label{urho1}
\end{equation}

The second error part in $u,u'$ can be derived from~(\ref{step}), substituting the residual
gradient $\bmath g$ that appears after the last refining iteration:
\begin{equation}
\left( \sigma_u^{(2)}\atop \sigma_{u'}^{(2)} \right) = - \mathbmss H_{\rm rsd}^{-1} \bmath g_{\rm rsd}.
\label{uu2}
\end{equation}
From~(\ref{rhoTayl}), this will imply the following error in $\rho$:
\begin{equation}
\sigma_\rho^{(2)} \simeq \frac{1}{2} \bmath g_{\rm rsd}^{\rm T} \mathbmss H_{\rm rsd}^{-1} \bmath g_{\rm rsd}.
\label{urho2}
\end{equation}

The third error source comes from possible round-off errors in $\bmath g$, which may
perturbate the computed position of the local minimum of $\rho$ via~(\ref{step}). Let us
assume that the numeric uncertainty of $\bmath g$ is $\sigma_g$, which has the meaning of a
typical length of the computed vector $\bmath g$ at the strict (algebraic) stationary point
of $\rho$, where $\bmath g$ must vanish. Then from~(\ref{step}) one may derive that
\begin{equation}
|\bmath d| \leq \frac{|\bmath g|}{|\lambda_{\min}|},
\end{equation}
where the $\lambda_{\min}$ is the minimum eigenvalue of $\mathbmss H$:
\begin{equation}
\frac{1}{|\lambda_{\min}|} = \frac{|\lambda_{\max}|}{|\det\mathbmss H|}.
\end{equation}
Hence, the corresponding uncertainty in $u,u'$ is estimated by
\begin{equation}
\sigma_{u,u'}^{(3)} \simeq \frac{|\lambda_{\max}|}{|\det\mathbmss H|}\sigma_g,
\label{uu3}
\end{equation}
The corresponding uncertainty for $\rho$ can be expressed using~(\ref{urho2}), but now
$\bmath g$ is basically an unknown random vector of average length $\sim \sigma_g$. We can
again apply the Rayleigh inequality to obtain
\begin{equation}
\sigma_\rho^{(3)} \simeq \frac{|\lambda_{\max}|}{2} \frac{\sigma_g^2}{|\det\mathbmss H|}.
\label{urho3}
\end{equation}
The quantity $\sigma_g$ appears more complicated and is derived below.

The fourth error source in $\rho$ appears when applying the first formula of~(\ref{vgH}).
If MOID is small then the relative error in $\rho$ increases due to the precision loss,
appearing because of the subtraction of close vectors $\bmath r$ and $\bmath r'$. If these
vectors have relative `storage' errors about $\nu \epsilon$ then their absolute numeric
errors are $\sigma_{\bmath r} \sim \nu\epsilon r$ and $\sigma_{\bmath r'} \sim \nu\epsilon
r'$. Hence, the inferred cumulative uncertainty of the difference is about their quadrature
sum:
\begin{equation}
\sigma_{\bmath r-\bmath r'} \simeq \nu\epsilon \sqrt{r^2+{r'}^2}.
\label{sdiff}
\end{equation}

Let us compute the inferred uncertainty in $\rho$ using the so-called delta method. Since
$2aa'\rho=(\bmath r-\bmath r')^2$ then the error $\Delta\rho$ appearing due to a small
perturbation $\Delta_{\bmath r-\bmath r'}$ is
\begin{equation}
2aa' \Delta\rho = 2(\bmath r-\bmath r') \Delta_{\bmath r-\bmath r'} + \Delta_{\bmath r-\bmath r'}^2.
\end{equation}
By replacing the terms above with their uncertainties or with absolute values of vectors,
and assuming that different terms are always added in the worst-case fashion, the final
error component in $\rho$ may be estimated as follows:
\begin{equation}
\sigma_\rho^{(4)} \simeq 2\rho \frac{\sigma_{\bmath r-\bmath r'}}{\sqrt{2aa'}} + \frac{\sigma_{\bmath r-\bmath r'}^2}{2aa'}.
\label{urho4}
\end{equation}
Notice that we cannot in general neglect the last term in~(\ref{urho4}), because it may
appear significant if $\rho$ is small.

Now we can also estimate the numeric uncertainty of $\bmath g$ by computing its finite
difference from~(\ref{vgH}), analogously to $\Delta\rho$:
\begin{equation}
2aa' \Delta \bmath g \simeq \left(\Delta_{\bmath r-\bmath r'}\bmath r_u + (\bmath r-\bmath r')\Delta \bmath r_u \atop \Delta_{\bmath r-\bmath r'}\bmath r'_{u'} + (\bmath r-\bmath r')\Delta \bmath r'_{u'}\right).
\end{equation}
By applying the same approach as for $\Delta\rho$, we may obtain the following for the
uncertainties in $\bmath g$:
\begin{equation}
2aa' \left(\sigma_{\bmath g_u}\atop \sigma_{\bmath g_{u'}}\right) \simeq \left(\sigma_{\bmath r-\bmath r'} r_u + \rho\sqrt{2aa'} \sigma_{\bmath r_u} \atop \sigma_{\bmath r-\bmath r'} r'_{u'} + \rho\sqrt{2aa'}\sigma_{\bmath r'_{u'}} \right).
\end{equation}
Since $\sigma_g^2 = \sigma_{\bmath g_u}^2 + \sigma_{\bmath g_{u'}}^2$, and $\sigma_{\bmath
r_u} \sim \nu\epsilon r_u$, and using~(\ref{sdiff}), one can obtain
\begin{equation}
\sigma_g \simeq \frac{\nu\epsilon}{2aa'}\left(|\bmath r-\bmath r'|+\sqrt{r^2+r'^2}\right) \sqrt{r_u^2+{r'_{u'}}^2}.
\end{equation}
But now we can clearly see that the term $|\bmath r-\bmath r'|$ is either small or of the
same order as $\sqrt{r^2+r'^2}$, so to simplify the formula we can simply neglect it and
leave only the latter term. We therefore put:
\begin{equation}
\sigma_g \simeq \frac{\nu\epsilon}{2aa'} \sqrt{r^2+r'^2} \sqrt{r_u^2+{r'_{u'}}^2}.
\end{equation}
This should be substituted to the formula~(\ref{urho3}) above.

Finally, summing up all four error components in $\rho$ yields the cumulative uncertainty
\begin{equation}
\sigma_\rho \sim \sigma_\rho^{(1)}+\sigma_\rho^{(2)}+\sigma_\rho^{(3)}+\sigma_\rho^{(4)}.
\end{equation}
Of course, some of these error terms may often become negligible, but it is difficult to
predict in advance which terms would dominate in this sum. Since either term may appear
large in certain conditions, we need to preserve all of them for the sake of reliability.

After that, an indicative uncertainty for the $\mathrm{MOID}=\sqrt{2aa'\rho}$ is
approximated by using the delta method as
\begin{equation}
\sigma_{\mathrm{MOID}} \sim \frac{aa'\sigma_\rho}{\mathrm{MOID}}.
\end{equation}
This formula is valid only if $\rho$ is not close to zero (compared to $\sigma_\rho$).
Otherwise, the MOID uncertainty is
\begin{equation}
\sigma_{\mathrm{MOID}} \sim \sqrt{2aa'\sigma_\rho}.
\end{equation}
The two latter formulae can be combined into a single approximate one:
\begin{equation}
\sigma_{\mathrm{MOID}} \sim \frac{aa'\sigma_\rho}{\sqrt{\mathrm{MOID}^2 + aa'\sigma_\rho/2}}.
\end{equation}

\subsection{Self-testing numerical reliability}
Finally, our algorithm includes a self-diagnostic test that verifies the following
post-conditions:
\begin{enumerate}
\item All roots that passed~(\ref{thr}) must comply with the requested least accuracy:
$\nu\varepsilon_z<\delta_{\max}$.
\item The minimum of $\Delta_z$ among all the roots that failed~(\ref{thr}) must exceed
$10$, meaning that there is no other suspicious root candidates. That is, the families
$|z|=1$ and $|z|\neq 1$ must be separated by a clear gap.
\item The number of roots that passed~(\ref{thr}) must be even and greater than four
(necessary algebraic conditions following from the theory).
\item After the 2D refining, the Hessian $\mathbmss H_{\rm rsd}$ is strictly
positive-definite, so we indeed are at a local minimum (rather than maximum or saddle
point).
\item On the 2D refining stage, the total cumulative change in $u$ satisfies the condition
$|\Delta u| < \delta_{\max}$. In some part this condition duplicates (i), ensuring the
initial approximation of the corresponding root did not appear to have inacceptable
accuracy. But it also ensures that the 2D refining did not switch us to a completely
different root of $g(u)$ (another stationary point of $\rho$). We pay no attention to $u'$
here, because it is always derived from $u$ using~(\ref{us}), so its numeric error, even if
large, is not indicative regarding the selection of a correct root of $g(u)$.
\end{enumerate}
If some of these conditions are broken, the algorithm sets a warning flag. Receiving such a
signal, the results should be considered unreliable. In practice this is a very seldom case
(see next section), but still possible. Then the following sequence can be done to verify
the results: (i) run the same algorithm on the same orbits $\mathcal E$ and $\mathcal E'$,
but swap them with each other; (ii) if failed again, run the same algorithm using the {\sc
long double} precision instead of {\sc double}; (iii) run the {\sc long double} computation
on swapped orbits; (iv) if everything failed, invoke an alternative MOID algorithm, e.g.
the one from Sect.~\ref{sec_add}.

We notice that since the task is solved asymmetrically, the algorithm may yield slightly
different results when computing $\mathrm{MOID}(\mathcal E,\mathcal E')$ and
$\mathrm{MOID}(\mathcal E',\mathcal E)$. If the orbital configuration does not imply
degeneracies, both computations should offer the same MOID value, within the reported
uncertainties. If they differ unacceptably, this can serve as an additional indicator that
something went wrong.

However, this notice does not apply to the estimated MOID uncertainty. This uncertainty can
appear different when computing $\mathrm{MOID}(\mathcal E,\mathcal E')$ and
$\mathrm{MOID}(\mathcal E',\mathcal E)$, because the polynomials $g(u)$ and $g(u')$ may
have (and usually do have) different algebraic properties. In fact, if the goal is accuracy
rather than speed, one may always compute the MOID in the both directions, $\mathcal E \to
\mathcal E'$ and $\mathcal E' \to \mathcal E$, selecting the value offering better
accuracy.

\section{On the choice of error tolerances}
\label{sec_tols}
The algorithm involves three main parameters related to the error control: $\delta_{\min}$,
$\delta_{\max}$, and $\nu$.

The primary error tolerance is $\delta_{\min}$. It controls the resulting accuracy of the
roots $z_k$, and of the eccentric anomalies $u,u'$, but not of the MOID itself and not even
of the adimensional function $\rho$. By setting $\delta_{\min}$ to a larger or smaller
value we may obtain less or more accurate result (in terms of $u,u'$). We can set
$\delta_{\min}=0$, meaning to seek maximum precision possible with the hardware (though
this probably requires to use the {\sc long double} arithmetic, see below).

The auxiliary error tolerance $\delta_{\max}$ does not actually control the accuracy of the
results. Setting a smaller $\delta_{\max}$ won't result in a more numerically accurate MOID
estimate. This parameter has two aspects: (i) it is used in the root-finding part to
control the initial `burn-in' stage of the Newton scheme and (ii) it is used as an
indicative threshold to separate numerically `reliable' cases from `unreliable' ones.

Therefore, the common sense requires that $\delta_{\max}$ must be greater (preferably,
significantly greater) than $\delta_{\min}$. Forcing $\delta_{\max}$ too small may result
in the following undesired effects. First, the Newton root-finding scheme may drastically
slow down, because its `burn-in' stage does not expect that it may reach the machine
precision and may otherwise iterate the roots until the internal iteration limit. The
output precision would then be worse than $\delta_{\max}$ anyway. Secondly, too small
$\delta_{\max}$ may trigger an unnecessary increase of the number of the unreliability
warnings.

Concerning the unreliability warnings, the practical value of $\delta_{\max}$ can be
selected based on the observational uncertainty of orbital elements (relative uncertainty
in terms of $a,a'$ or absolute one in terms of the angular elements). This input
uncertainty is typically considerably larger than the machine precision. With so-selected
$\delta_{\max}$, the warning flag would indicate that the numeric accuracy of $z_k$ might
be worse than the errors inferred by input observational uncertainties. Then the
intermediate (inferred by $z_k$) numeric uncertainty of $u$ and $u'$ may appear larger than
what we can trigger by varying the orbital elements within their error boxes. In this case
the warning is physically reasonable. But if the numeric uncertainty always remains below
the observational one then signaling any warning does not make sense.

In any case, it does not make sense to set $\delta_{\max}$ too much below the physical
sizes of the objects (relative to $\sim \max(a,a')$). For the Main Belt, this is $\sim
10^{-11}$~AU corresponding to the smallest known asteroids of $\sim 1$~m in size.

If the uncertainty of orbital elements is unknown or irrelevant than the good choice of
$\delta_{\max}$ is $\sqrt{\delta_{\min}}$, which follows from the properties of the
Newton-Raphson method. In such a case, for each root $z_k$ we need to make just one or two
iterations after the `burn-in' part of the Newton scheme. This is because the number of
accurate digits is roughly doubled after each Newtonian iteration. For example, if the
accuracy of $\sqrt\epsilon$ has been reached, on the next iteration we will likely have
$\epsilon$.

As to the last control parameter, $\nu$, it may be used to manually scale up all the error
assessments. So far in our tests we did not find practical reasons to set it to something
different from $\nu=1$. But whenever necessary, it can be used to disentangle the value
$\delta_{\max}$ used by the Newtonian root-finding scheme from the threshold used in the
error control part. Since $\nu$ scales all the error estimates up, its effect is equivalent
to reducing the error threshold from $\delta_{\max}$ to $\delta_{\max}/\nu$, but the Newton
scheme is always using $\delta_{\max}$ and ignores the scale factor $\nu$.

Summarizing, it appears that in the general case it is reasonable to set $\delta_{\min}$
about the machine epsilon, $\delta_{\max}\sim\sqrt{\delta_{\min}}$, and to select such
$\nu$ that $\delta_{\max}/\nu$ is about the physically justified MOID uncertainty (relative
to $\sim \max(a,a')$).

\section{Practical validation and performance benchmarks}
\label{sec_perf}
We tested our algorithm on the first $10000$ numbered asteroids from the Main Belt,
implying $\sim 10^8$ orbit pairs. The orbital elements were taken from the catalog
\texttt{astorb.dat} of the Lowell observatory.\footnote{See url
ftp://ftp.lowell.edu/pub/elgb/astorb.html.}

Our algorithm succeeded nearly always. When using the standard {\sc double} floating-point
arithmetic, the self-test conditions listed above were failed only once per $25000$ orbit
pairs. In case of such a warning the first attempt was to rerun the same algorithm
interchanging the orbits $\mathcal E$ and $\mathcal E'$. Since the method treats orbits
asymmetrically, this usually helps. Double-warnings occured in our test once per $2.5\times
10^6$ orbit pairs.

We note that if the algorithms returns a bad self-diagnostic flag, this does not yet mean
that it failed to compute the MOID and the result is necessarily wrong or just absent. One
of the reasons for a warning is that some root $z_k$ (not even necessarily related to the
global minimum) is worse than the required least accuracy $\delta_{\max}$. But worse does
not mean necessarily useless. This just means that the result needs an attention and
probably needs a more detailed investigation using different other methods to confirm or
refine the results. Occurences when the resulting MOID appears entirely wrong and has
inacceptable accuracy, represent only a small fraction of all those cases when the warning
was reported.

\begin{figure}[!t]
\includegraphics[width=\linewidth]{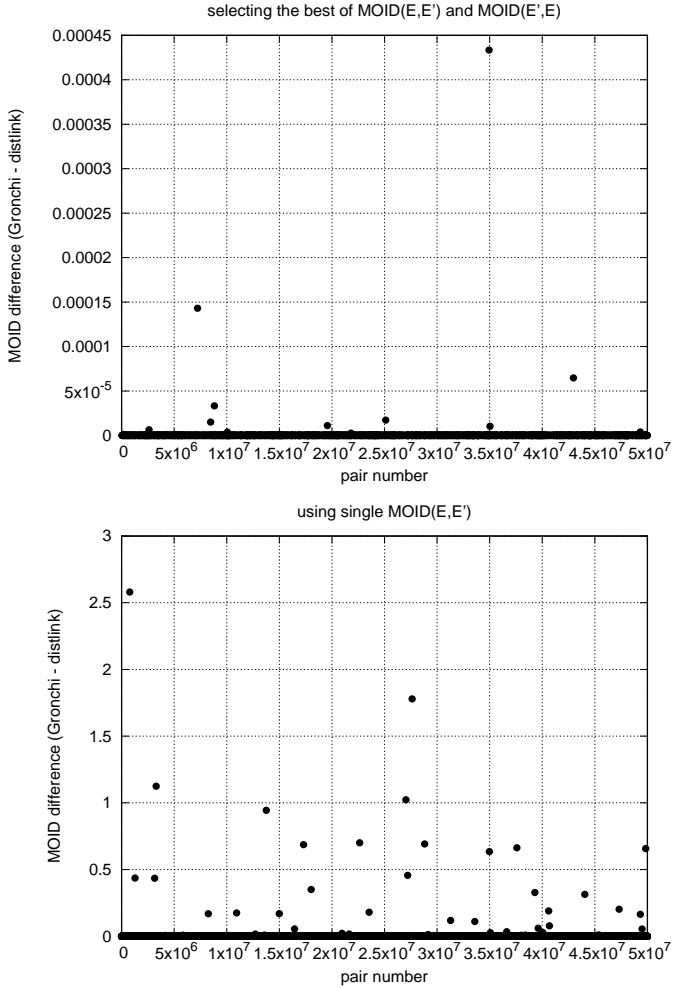}
\caption{The difference between MOID values computed by the Gronchi's code and by our new
algorithm (labelled as {\sc distlink}). Top: comparing results improved by orbits
interchanging and selecting the best MOID of the two. Bottom: using only a single MOID
value for comparison. To reduce the figure file size, we removed from the both graphs all
differences below $10^{-13}$~AU in absolute magnitude. In these conditions, no points were
revealed below the abscissa, i.e. {\sc distlink} always provided smaller MOID value than
the Gronchi's code. See text for a detailed explanation.}
\label{fig_Gtest}
\end{figure}

We also tested the Gronchi FORTRAN code in the same setting. We found only two orbit pairs
for which it failed with a error and no-result, and swapping the orbits did not help. A
single-fail case, when the first attempt to compute MOID failed, but swapping the orbits
did help, occured once per $\sim 3\times 10^5$ MOID computations. For the majority of orbit
pairs this algorithm gave some numeric MOID value at least, but in itself this does
guarantees that all these values are accurate.

We provide a comparison of our new algorithm with the Gronchi code in Fig.~\ref{fig_Gtest}.
We compute the differences of the Gronchi MOID minus the MOID obtained by our new algorithm
in two settings. In the first case, we run both algorithms for each orbit pair twice, to
compute $\mathrm{MOID}(\mathcal E, \mathcal E')$ and $\mathrm{MOID}(\mathcal E', \mathcal
E)$. With the Gronchi code, we select the minimum MOID between the two, and for our new
code we select the best-accuracy MOID. If Gronchi algorithm failed with no-result in one of
the two runs, the corresponding MOID value was ignored, and only the other one was used. If
both values of the MOID obtained by Gronchi's algorithm were failed, this orbit pair itself
was ignored. Additionally, if our new algorithm reported a warning, we either ignored this
MOID in favour of the other value, or invoked the fallback algorithm from
Sect.~\ref{sec_add}, if that second MOID estimate appeared unreliable too. The MOID
difference between the Gronchi code and new algorithm was then plotted in the top panel of
Fig.~\ref{fig_Gtest}. In the second setting, we plainly performed just a single MOID
computation for each orbit pair without orbit interchange, either using the Gronchi code or
our new algorithm. Orbit pairs for which Gronchi code failed or our algorithm reported a
warning, were ignored and removed. The corresponding MOID difference is plotted in the
bottom panel of Fig.~\ref{fig_Gtest}.

We may see that there are multiple occurences when Gronchi code obtained clearly
overestimated MOID value (i.e., it missed the true global minimum). But all the cases, in
which Gronchi algorithm produced smaller MOID than our library, correspond to the MOID
difference of $\sim 10^{-13}$~AU at most, with $\sim 10^{-16}$~AU in average. So all these
occurences look like some remaining round-off errors (possibly even in the Gronchi code
rather than in {\sc distlink}). Therefore, we did not find an occurence in which {\sc
distlink} would yield clearly wrong MOID value without setting the unreliability flag.

\begin{figure}[!t]
\includegraphics[width=0.49\textwidth]{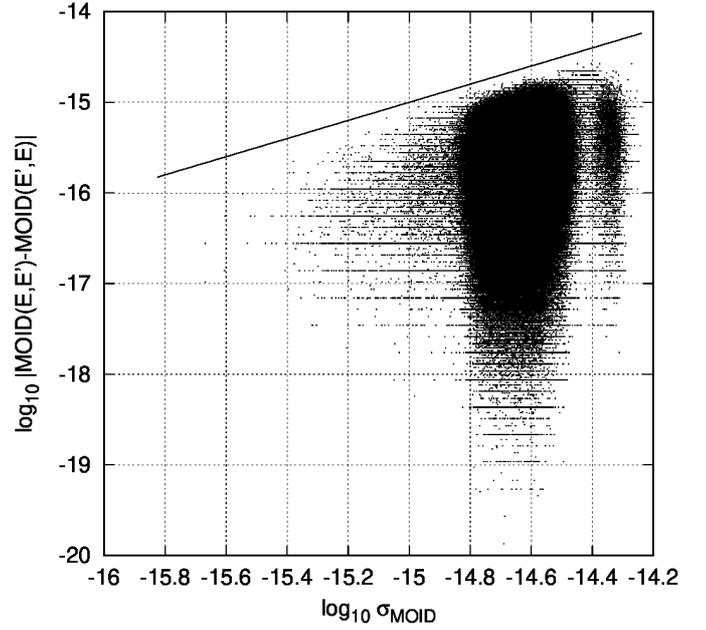}
\caption{Distribution of the estimated uncertainties
$\sigma_\mathrm{MOID}$ versus an empiric error measure $\left|\mathrm{MOID}(\mathcal E,
\mathcal E') - \mathrm{MOID}(\mathcal E', \mathcal E)\right|$, for the test case of $10^8$
orbital pairs (see text). The inclined line labels the main diagonal (abscissa equals
ordinate). All simulated dots falled below this line. The computations were done in the
{\sc double} floating-point arithmetic, AMD FX configuration.}
\label{fig_testerr}
\end{figure}

In Fig.~\ref{fig_testerr} we compare the quadrature sum of the reported MOID uncertainties,
$\sigma_\mathrm{MOID}=\sqrt{\sigma_\mathrm{MOID(\mathcal E,\mathcal E')}^2 +
\sigma_\mathrm{MOID(\mathcal E',\mathcal E)}^2}$, with the difference
$|\mathrm{MOID}(\mathcal E,\mathcal E')-\mathrm{MOID}(\mathcal E',\mathcal E)|$ that can be
deemed as an empiric estimate of the actual MOID error. We may conclude that our algorithm
provides rather safe and realistic assessment of numeric errors, intentionally a bit
pessimistic. We did not met a case with the empiric error exceeding the predicted
uncertainty.

\begin{figure*}[!t]
\includegraphics[width=0.49\textwidth]{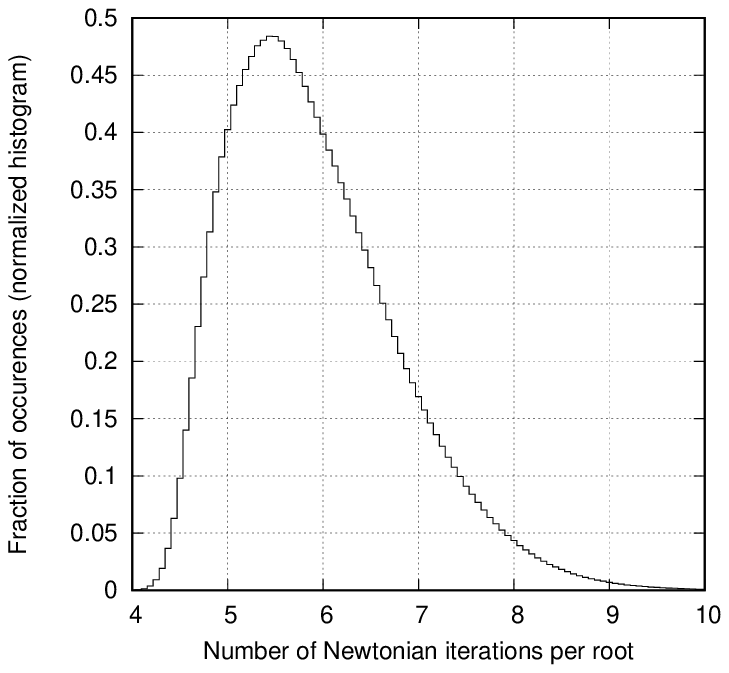}
\includegraphics[width=0.49\textwidth]{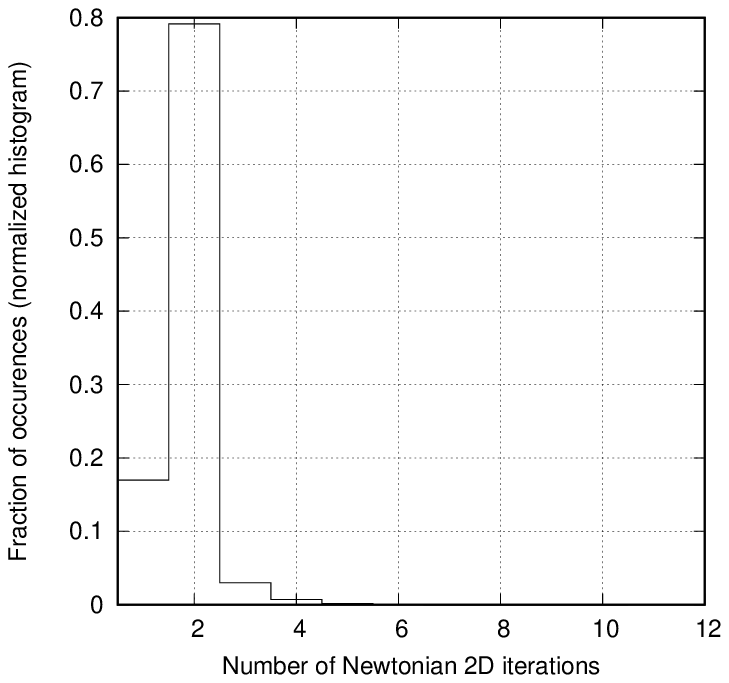}
\caption{Histogram for the number of Newtonian iterations spent per root (left) and of 2D
Newtonian iterations made on the final refine stage (right). The histograms were normalized
by the bin width, so they actually render the probability density function for the quantity
labelled in the abscissa. The computations were done in the {\sc double} floating-point
arithmetic, AMD FX configuration.}
\label{fig_testiter}
\end{figure*}

From Fig.~\ref{fig_testiter} one can see that we spend, in average, about $n=5-6$ Newtonian
iterations per each root. One way how we may further increse the speed of computation is to
reduce this number. However, this number is already quite small, so there is no much room
to significantly reduce it. On the refining stage the algorithm usually performs just one
or two 2D Newtonian iterations in the plane $(u,u')$. The fraction of occurences when three
or more refining iterations are made is very small and further decreases quickly. The
maximum number of refining iterations made in this test was $7$.

\begin{table*}
\caption{Performance tests on the first $10000$ asteroids from the Main Belt: average CPU time per MOID.}
\begin{center}
\begin{tabular}{lcccc}
\hline
Hardware           & \multicolumn{2}{c}{{\sc double} arithmetic} & \multicolumn{2}{c}{{\sc long double} arithmetic} \\
                   & {\sc distlink}& Gronchi code  & {\sc distlink}& Gronchi code   \\
                   & (fast alg.)   &               & (fast alg.)   &                \\
\hline
Intel Core i7      & $24$~$\mu$s   & $36$~$\mu$s   &  $77$~$\mu$s  &  NA   \\
Supermicro \& Xeon & $31$~$\mu$s   & $61$~$\mu$s   & $100$~$\mu$s  &  NA   \\
AMD FX             & $44$~$\mu$s   & $70$~$\mu$s   & $357$~$\mu$s  &  NA   \\
\hline
\end{tabular}
\end{center}
\label{tab_bench}
\end{table*}

In Table~\ref{tab_bench}, we present our performance benchmarks for this test application.
They were done for the following hardware: (i) Intel Core i7-6700K at $4.0$~GHz, (ii)
Supermicro server with Intel Xeon CPU E5-2630 at $2.4$~GHz, and (iii) AMD 990FX chipset
with AMD FX-9590 CPU at $4.4$~GHz. The second configuration is rather similar to one of
those used by \citet{Hedo18}.

We used either 80-bit {\sc long double} floating-point arithmetic or the 64-bit {\sc
double} one, and requested the desired accuracy of $2\epsilon$: $\delta_{\min}\sim
2.2\times 10^{-19}$ or $\delta_{\min}\sim 4.4\times 10^{-16}$, respectively. We did not use
$\delta_{\min}=0$, because in the {\sc double} case many CPUs hiddenly perform much of the
local computation in {\sc long double} precision instead of the requested {\sc double}.
Newtonian iterations are then continued to this undesiredly increased level of precision,
if $\delta_{\min}=0$, thus introducing an unnecessary minor slowdown. The least required
accuracy $\delta_{\max}$ was set to $\sqrt\epsilon$ in all of the tests.

All the code was compiled with {\sc GCC} (either \texttt{g++} or \texttt{gfortran}) and
optimized for the local CPU architecture (\texttt{-O3 -march=native -mfpmath=sse}). The
Gronchi primary computing subroutine \texttt{compute\_critical\_points\_shift()} was called
from our C++ program natively, i.e. without any intermediary file IO wrapping that would be
necessary if we used the main program \texttt{CP\_comp.x} from the Gronchi package.

To accurately measure the time spent purely inside just the MOID computation, and not on
the file IO or other algorithmic decorations around it, we always performed three
independent runs on the test catalog: (i) an `empty' run without any calls to any MOID
algorithm, only iteration over the catalog; (ii) computation of all MOIDs using the
algorithm of this paper, without writing results to a file; (iii) same for the Gronchi
algorithm. The time differences (ii)-(i) or (iii)-(i) gave us the CPU time spent only
inside the MOID computation. We never included the CPU time spent in the kernel mode. We
assume this system time likely refers to some memory pages manipulation or other similar
CPU activity that appears when the program iteratively accesses data from a big catalog. In
any case, this system time would be just a minor addition to the result ($\sim 1-2$ per
cent at most).

The reader may notice that the hardware can generate huge performance differences, not
necessarily owing to just the CPU frequency. Moreover, even the performance on the same AMD
machine differs drastically between the {\sc double} and {\sc long double} tests. This
puzzling difference appears mainly due to slow 80-bit floating-point arithmetic on AMD, not
because of e.g. different number of Newtonian iterations per root (which appeared almost
the same in all our tests, $5$--$6$ iterations per root).

We conclude that our algorithm looks quite competitive and probably even outperforming the
benchmarks obtained by \citet{Hedo18} for their set of tested algorithms ($60$--$80$~$\mu$s
per orbit pair on a Supermicro/Xeon hardware). They used {\sc double} precision rather than
{\sc long double} one.

Therefore, our algorithm possibly pretends to be the fastest one available to date, or at
least it belongs to the family of the fastest ones. In the majority of cases it yields
considerably more accurate and reliable results, usually close to the machine precision,
and its accuracy may seriously degrade only in extraordinary rare nearly degenerate cases,
which are objectively hard to process.

\section{Additional tools}
\label{sec_add}
Our main algorithm based on determining the roots of $g(u)$ is fast but might become
vulnerable in the rare cases of lost roots. Whenever it signals a warning, alternative
algorithms should be used, trading computing speed for better numeric resistance with
respect to degeneracies.

In addition to the basic 0D method based on $g(u)$ root-finding, our library implements a
``fallback'' algorithm of the 1D type, based on the brute force-like minimization of
$\tilde\rho(u)$. This method is numerically reliable thanks to its simplicity, and its slow
speed is not a big disadvantage, because it needs to be run only if the basic fast method
failed. In our benchmarking test it appeared $\sim 6$ times or $\sim 4$ times slower than
our fast algorithm or the Gronchi code, respectively. But this is likely sensitive to its
input parameters.

First of all, the algorithm scans only a restricted range in the $u$ variable, discarding
the values where the MOID cannot be attained. The requied $u$ range is determined as
follows. Using e.g. formulae from \citep{KholshVas99lc}, compute the minimum internodal
distance $d_\Omega$. Since MOID is usually attained near the orbital nodes, this quantity
and its corresponding orbital positions already provide rather good approximation to the
MOID. Then consider two planes parallel to the orbit $\mathcal E'$, and separated from it
by $\pm d_\Omega$. We need to scan only those pieces of orbit $\mathcal E$ that lie between
these planes, i.e. lie within $\pm d_\Omega$ band from the $\mathcal E'$ plane. The points
on $\mathcal E$ outside of this set are necessarily more distant from $\mathcal E$ than
$d_\Omega$, so the MOID cannot be attained there. This trick often reduces the $u$ range
dramatically. This optimization was inspired by the discussion given in \citep{Hedo18}. The
detailed formulae for the reduced range of $u$ are given in~\ref{sec_urange}.

Moreover, this algorithm automatically selects the optimal orbits order $(\mathcal
E,\mathcal E')$ or $(\mathcal E',\mathcal E)$ to have a smaller angular range to scan. In
the case if the cumulative ranges appear equal (e.g. if we occasionally have the
full-circle $[0,2\pi]$ in both cases) then the user-supplied order is preserved.

\begin{figure}[!t]
\includegraphics[width=\linewidth]{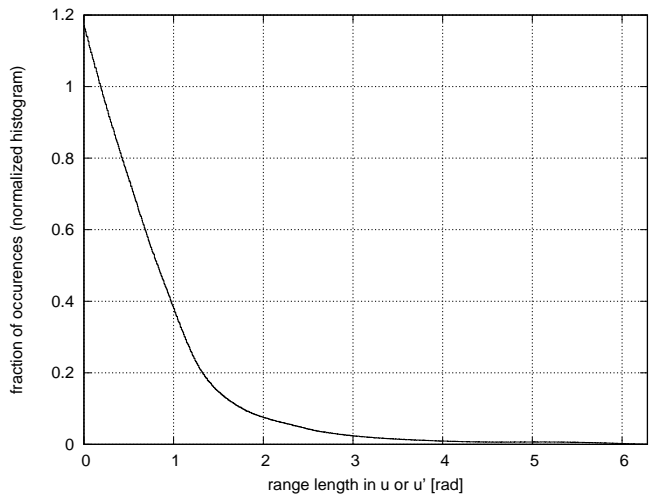}
\caption{The distribution density of the reduced angular range in $u$, as obatined for
$\sim 10^8$ asteroid pairs.}
\label{fig_urange}
\end{figure}

The efficiency of this approach is demonstrated in Fig.~\ref{fig_urange}, where we plot the
distribution density for the total range length obtained, as computed for our test case of
$10^4\times 10^4$ asteroid pairs. The fraction of the cases in which this range could not
be reduced at all (remained at $[0,2\pi]$) is only $\sim 2\%$, and in the majority of
occurences it could be reduced to something well below $1$~rad. The efficiency of the
reduction increases if MOID is small. Then the total scan range may be reduced to just a
few degrees.

The minimization of $\tilde\rho(u)$ is based on subsequent sectioning of the initial
angular range for $u$. The user can set an arbitrary sequence of integer numbers
$n_1,n_2,n_3,\ldots,n_p$ that define how many segments are considered at different stages.
The initial angular range is partitioned into $n_1$ equal segments separated by the
equidistant nodes $u_k$, and the node with a minimum $\tilde\rho(u_k)$ is determined. We
note that the input parameter $n_1$ is always interpreted as if it corresponded to the
entire $[0,2\pi]$ range, even if the actual scan range is reduced as described above. The
segment length on the first step is normally set to $h_1=2\pi/n_1$ regardless of the scan
range, unless this scan range is itself smaller than $h_1$. On the second stage, the
segment $[u_{k-1},u_{k+1}]$ surrounding the minimum $u_k$ is considered. It is sectioned
into $n_2$ equal segments, and the node corresponding to the minimum $\tilde\rho(u_k)$ is
determined again. On the third stage, the segment $[u_{k-1},u_{k+1}]$ is sectioned into
$n_3$ smaller segments, and so on. On the $k$th stage, the length of the segment between
subsequent nodes is reduced by the factor $2/n_k$, so only $n_k\geq 3$ are meaningful.
Starting from the stage number $p$, the segments are always partitioned into $n_p$
subsegments, until the global minimum of $\tilde\rho(u)$ is located with a desired accuracy
in $u$ and $\rho$. It is recommended to set $n_1$ large enough, e.g. $\sim 1000$, in order
to sample the objective function with its potentially intricate variations densely enough,
whereas the last $n_p$ can be set to $4$, meaning the bisection.

We notice that this method was not designed for a standalone use. It was not even supposed
to be either more reliable or more accurate in general than our primary fast method. It was
supposed to provide just yet another alternative in rare special cases when the primary
method did not appear convincingly reliable. Its practical reliability depends on the input
parameters very much: too small $n_1$ may lead to frequent loosing of local minima in
$\tilde\rho(u)$, if they are narrow. Hence we may miss the correct global minimum
sometimes. But this effect can be always suppressed by selecting a larger $n_1$. In our
tests, with $n_1=1000$ this algorithm generated one wrong MOID per $\sim 3000$ trials, so
it is not recommended for a general sole use. This could be improved by implementing an
adaptive sampling in the $u$ variable, e.g. depending on the derivative $|\tilde\rho'|$,
but we did not plan to go that far with this method. We note that narrow local minima of
$\rho$ are, informally speaking, in some sense antagonistic to almost-multiple critical
points, so this fallback algorithm is vulnerable to rather different conditions than our
primary fast method. Therefore it can serve as a good complement to the latter.

Also, we include in the library several fast tools that may appear useful whenever we need
to actually compute the MOID only for those object that are close to an orbital
intersection. These tools may help to eliminate most of the orbit pairs from the
processing. The first one represents an obvious pericenter--apocenter test: $\mathrm{MOID}
\geq a(1-e)-a'(1+e')$, and $\mathrm{MOID} \geq a'(1-e')-a(1+e)$. If any of these quantities
appeared positive and above some upper threshold $\mathrm{MOID}_{\max}$ then surely
$\mathrm{MOID}>\mathrm{MOID}_{\max}$, and one may immediately discard such orbital pair
from the detailed analysis.

Our library also includes functions for computing the so-called linking coefficients
introduced by \citet{KholshVas99lc}. The linking coefficients are functions of two orbits
that have the dimension of squared distance, like $|\bmath r-\bmath r'|^2=2aa'\rho$, and
they are invariable with respect to rotations in $\mathbb R^3$. \citet{KholshVas99lc}
introduced three linking coefficients that should be selected depending on whether the
orbits are (nearly) coplanar or not. See all the necessary formulae and discussion in that
work.

For our goals it might be important that at least one of these linking coefficients, $l_1 =
d_1 d_2$, can be used as an upper limit on the MOID. It represents a signed product of two
internodal distances from~(\ref{ind}), so the squared MOID can never be larger than
$|l_1|$. This allows us to limit the MOID from another side, contrary to the
pericenter-apocenter test. Moreover, based on $l_1$ we introduce yet another linking
coefficient defined as
\begin{equation}
l_1' = \min\left(|d_1|,|d_2|\right)^2 \sign l_1.
\end{equation}
This modified $l_1$ provides even tighter upper limit on the squared MOID, but still
preserves the sign that indicates the orbital linkage in the same way as $l_1$ did.

It is important for us that all linking coefficients are computed very quickly in
comparison to any MOID algorithm, because they are expressed by simple elementary formulae.

The linking coefficients were named so because their original purpose was to indicate
whether two orbits are topologically linked like two rings in a chain, or not. The
intermediate case between these two is an intersection, when the linking coefficient
vanishes together with the MOID. Therefore, these indicators can be potentially used as
computationally cheap surrogates of the MOID. But in addition to measuring the closeness of
two orbits to an intersection, linking coefficients carry information about their
topological configuration. Also, these quantities can be used to track the time evolution
of the mutual topology of perturbed non-Keplerian orbits, for example to locate the moment
of their intersection without computing the MOID.

\section{Further development and plans}
Yet another possible way to extend our library is to implement the method by
\citet{Baluev05} for computing the MOID between general confocal unperturbed orbits,
including hyperbolic and parabolic ones. This task can be also reduced to finding real
roots of a polynomial similar to $\mathcal P(z)$.

In a future work we plan to provide statistical results of applying this algorithm to the
Main Belt asteroids, also including the comparison of the MOID with linking coefficients
and other indicators of orbital closeness.

\section*{Acknowledgements}
This work was supported by the Russian Science Foundation grant no.~18-12-00050. We express
gratitude to the anonymous reviewers for the fruitful comments and useful suggestions on
the manuscript.

\appendix

\section{Reducing the scan range for the eccentric anomaly}
\label{sec_urange}
 Let us introduce $\bmath R' = \bmath P' \times
\bmath Q'$, which is a unit vector orthogonal to the orbital plane of $\mathcal E'$. The
vectors $\bmath P'$, $\bmath Q'$, $\bmath R'$ form an orthonormal basis in $\mathbb R^3$.
Then from~(\ref{rvec}) let us compute the dot-product
\begin{equation}
(\bmath r - \bmath r') \bmath R' = a PR' (\cos u - e) + a SR' \sin u,
\end{equation}
which represents a projection of the distance vector $\bmath r - \bmath r'$ on the basis
vector $\bmath R'$. Note that the dot-product $\bmath r' \bmath R'$ is always zero. Now, we
need this distance projection to be within $\pm d_\Omega$ from zero, because otherwise the
absolute distance can be only larger than $d_\Omega$. This yields two inequality
constraints
\begin{equation}
e PR' - \frac{d_\Omega}{a} \leq PR' \cos u + SR' \sin u \leq e PR' + \frac{d_\Omega}{a},
\label{ineq}
\end{equation}
implying an elementary trigonometric equation that can be solved via arcsines.

The final set of computing formulae can be expressed as follows. Let us introduce the
vector
\begin{equation}
\bmath W = \bmath R \times \bmath R', \quad W=|\bmath W|=\sin I,
\end{equation}
which is directed to the ascending node of $\mathcal E'$ assuming reference $\mathcal E$.
The angle $I$ is the mutual inclination between the orbits. Then determine the angle
$\theta$ from
\begin{equation}
 \cos\theta = (PW)/W, \quad \sin\theta = (QW)/W.
\end{equation}
It represents the true anomaly on $\mathcal E$, where that ascending node is located.
Basically, $\theta$ is the angle between $\bmath P$ and $\bmath W$, counted positive in the
direction of $\bmath Q$. The location on the other orbit $\theta'$ can be determined in a
similar way. Explicit formula for the scalar product $PW$ is given in
\citep{KholshVas99lc} via orbital elements, though we prefer to multiply the vectors
directly, using the following expression for $\bmath W$:
\begin{align}
\bmath W = \{&\cos i \sin i' \cos \Omega' - \sin i \cos i' \cos \Omega, \nonumber\\
&\cos i \sin i' \sin \Omega' - \sin i \cos i' \sin \Omega, \nonumber\\
&\sin i \sin i' \sin(\Omega'-\Omega)\, \}. \nonumber\\
\end{align}
After that let us compute
\begin{align}
d_1 &= \frac{p}{1+e\cos\theta} - \frac{p'}{1+e'\cos\theta'}, \nonumber\\
d_2 &= \frac{p}{1-e\cos\theta} - \frac{p'}{1-e'\cos\theta'}, \nonumber\\
d_\Omega &= \min(|d_1|,|d_2|),
\label{ind}
\end{align}
where $p$ and $p'$ are orbital parameters, $p=a(1-e^2)$.

Now, the inequalities~(\ref{ineq}) may be simplified if we decompose the vectors $\bmath W$
and $\bmath R'$ in the basis $\{\bmath P, \bmath Q, \bmath R\}$:
\begin{align}
\bmath W &= \{PW,\, QW,\, RW=0\,\}, \nonumber\\
\bmath R' &= \{PR',\, QR',\, RR'=\cos I\,\}.
\end{align}
Writing down the orthogonality condition between $\bmath W$ and $\bmath R'$ and the norm of
$\bmath R'$ in these coordinates, we have
\begin{align}
WR' &= PW\; PR' + QW\; QR' = 0, \nonumber\\
R'^2 &= 1 \implies PR'^2 + QR'^2 = W^2.
\end{align}
Therefore, we may set $PR' = \mp W\sin\theta$ and $QR' = \pm W\cos\theta$ in~(\ref{ineq}),
and the sign choice is not important here.

Finally, let us define the quantity $k\geq 0$ and the angle $\varphi$ from
\begin{align}
A^2 = 1 - e^2 \cos^2\theta, \quad k = \frac{d_\Omega}{a W A}, \nonumber\\
\sin\varphi = \frac{\sin\theta}{A}, \quad \cos\varphi = \sqrt{1-e^2}\, \frac{\cos\theta}{A},
\end{align}
and~(\ref{ineq}) becomes
\begin{equation}
e\sin\varphi - k \leq \sin(\varphi-u) \leq e \sin\varphi + k.
\end{equation}

In general, we have three types of solution for $u$.
\begin{enumerate}
\item If $|e\sin\varphi| < |1-k|$ and $k<1$ then we have two small segments for $u$ near the nodes,
defined as $[\varphi-\arcsin(e\sin\varphi+k), \varphi-\arcsin(e\sin\varphi-k)]$ and
$[\varphi+\pi+\arcsin(e\sin\varphi-k), \varphi+\pi+\arcsin(e\sin\varphi+k)]$;

\item If $|e\sin\varphi| < |1-k|$ and $k\geq 1$ then we have the entire circular range
$[0,2\pi]$ for $u$.

\item If $|e\sin\varphi| \geq |1-k|$ then there is just one big segment for $u$ that covers
angles roughly from one node to another, defined as either
$[\varphi+\arcsin(e\sin\varphi-k), \varphi+\pi-\arcsin(e\sin\varphi-k)]$, if
$\sin\varphi>0$, or $[\varphi-\arcsin(e\sin\varphi+k),
\varphi+\pi+\arcsin(e\sin\varphi+k)]$, if $\sin\varphi<0$;
\end{enumerate}
In practice, the first type of occurence is the most frequent one, so the speed improvement
is dramatic. Notice that for $W\to 0$ (coplanar orbits) the angle $\theta$ formally becomes
undefined, but this is not important because then $k\to\infty$ and we just obtain the
full-circle range $[0,2\pi]$ for $u$. So the degenerate case $W\approx 0$ is not a big
numeric issue in practice.

\bibliographystyle{model2-names}
\bibliography{distalg}

%% not want to use model2-names.bst.



%% \bibitem must have one of the following forms:
%%   \bibitem[Jones et al.(1990)]{key}...
%%   \bibitem[Jones et al.(1990)Jones, Baker, and Williams]{key}...
%%   \bibitem[Jones et al., 1990]{key}...
%%   \bibitem[\protect\citeauthoryear{Jones, Baker, and Williams}{Jones
%%       et al.}{1990}]{key}...
%%   \bibitem[\protect\citeauthoryear{Jones et al.}{1990}]{key}...
%%   \bibitem[\protect\astroncite{Jones et al.}{1990}]{key}...
%%   \bibitem[\protect\citename{Jones et al., }1990]{key}...
%%   \harvarditem[Jones et al.]{Jones, Baker, and Williams}{1990}{key}...
%%


% \end{thebibliography}

\end{document}